\documentclass[11pt,a4paper]{article}
\usepackage{jheppub_kim}

\usepackage{pdflscape}
\usepackage{amsmath}
\usepackage{amssymb}
\usepackage{dcolumn}
\usepackage{bm}
\usepackage{color}
\usepackage{epsfig}
\usepackage{amsfonts}
\usepackage{graphicx}
\usepackage{subfigure}
\usepackage{dcolumn}

\newcommand{\be}{\begin{equation}}
\newcommand{\ee}{\end{equation}}
\newcommand{\bea}{\begin{eqnarray}}
\newcommand{\eea}{\end{eqnarray}}

\def\be{\begin{equation}}
\def\ee{\end{equation}}
\def\bea{\begin{eqnarray}}
\def\eea{\end{eqnarray}}

\begin{document}

\title{The universe dominated by the extended Chaplygin gas}

\author[a]{E.O. Kahya,}

\author[b]{B. Pourhassan}

\affiliation[a]{Physics Department, Istanbul Technical University, Istanbul, Turkey}
\affiliation[b]{School of Physics, Damghan University, Damghan, Iran}

\emailAdd{eokahya@itu.edu.tr}
\emailAdd{bpourhassan@yahoo.com}

\abstract{In this paper, we consider a universe dominated by the extended Chaplygin gas which recently proposed as the last version of Chaplygin gas models. Here, we only consider the second order term which recovers quadratic barotropic fluid equation of state. The density perturbations analyzed in both relativistic and Newtonian regimes and show that the model is stable without any phase transition and critical point. We confirmed stability of the model using thermodynamics point of view.}


\maketitle

\section{Introduction}
Accelerated expansion of universe can be described by dark energy \cite{0603057,1212.4726}. In that case there are several dark energy candidates. One of the first and simplest model is the cosmological constant \cite{0004075,0207347} which has two important problems known as fine-tuning and coincidence. Therefore, alternative models were investigated such as quintessence \cite{PRD37(1988)3406,9708069,9807002,P94 1310.7167,P95 1311.4024,P97 1402.6821,P98 1403.3768} and k-essence \cite{0004134,0006373}.\\
Another attempt to solve the coincidence problem is interacting dark energy model \cite{9408025,9908023,0510628,0105479,P96 1402.2385,0303145,0411025,1312.1162,1306.2055,1302.1819}. The dark energy density is of the same order as the
dark matter energy density in the present universe, therefore one could imagine that there is
some connection between dark energy and dark matter.\\
The above models are based on modification of the scalar field and the fine-tuning of the potential. Instead, one can modify the equation of state to produce an exotic background fluid so-called Chaplygin gas. The pure Chaplygin gas emerged for the first time from the string theory point of view \cite{john1,john2}, then it is used to study classical solution of bosonic d-brane in ($d+1$,1) space-time \cite{P1 0003288}. One can begin from Nambu-Goto action in the light-cone gauge and solve the momentum constraint to obtain the Chaplygin gas equation. Then it was seen that different kinds of matter on the brane, e.g. Chaplygin gas, can be introduced  \cite{P2 0005011}. The Chaplygin gas equation of state is given by \cite{P4 0103004},
\begin{equation}\label{I2}
p_{CG}=-\frac{B}{\rho_{CG}},
\end{equation}
where $B$ is an arbitrary parameter which usually considered as a constant. In that case density perturbations in a universe dominated by the Chaplygin
gas has been studied by the Ref. \cite{GRG34(2002)53}.\\
In order for the universe to pass from a dust dominated epoch to a de
Sitter phase through an intermediate phase described as mixture of cosmological
constant and radiation the Chaplygin gas equation of state generalized to \cite{EPJC73(2013)2295,P16 0202064},
\begin{equation}\label{I3}
p_{GCG}=-\frac{B}{\rho_{GCG}^{\alpha}},
\end{equation}
where $\alpha$ and $B$ are free parameters. It is clear that $\alpha=1$ reproduces the pure Chaplygin gas model. This model is called the generalized Chaplygin gas model \cite{P17 0209395}. In the recent years the generalized Chaplygin gas was the subject of several cosmological and phenomenological studies. For example Brane-world cosmology and Brane-world inflation including the generalized Chaplygin gas has been studied \cite{P20 1006.1847,P21 1012.1050}. There is also possibility to consider viscosity in the generalized Chaplygin gas model \cite{P36 0511814,P37,P38,P39 1402.3669,P40 1305.6054}.\\
There is also a class of equations of state that interpolate between
standard fluids at high energy densities and generalized Chaplygin gas fluids at low energy densities which is called the modified Chaplygin
gas \cite{P47 0205140}. Viscous modified Chaplygin gas was introduced by the Ref. \cite{P54 0801.2008} and completed by the recent works \cite{P55 1301.2788,P56,P57 1401.8002}. Modified Chaplygin gas was introduced with the
following equation of state \cite{P83 1211.3518,MNRASoc397(2009)1935},
\begin{equation}\label{I5}
p_{MCG}=A\rho_{MCG}-\frac{B}{\rho_{MCG}^{\alpha}},
\end{equation}
where $A$, $\alpha$, and $B$ are parameters of the model. The case of $A = 0$ recovers generalized Chaplygin gas equation of state, and $A = 0$ together $\alpha = 1$ recovers the original Chaplygin gas equation of state.
The MCG equation of state has two parts, the first term gives an ordinary fluid obeying a linear barotropic equation of state, while there are some models with quadratic equation of state \cite{0512224,1309.5784,PRD80(2009)023008},
\begin{equation}\label{I1}
p = p_{0}+\omega_{1}\rho+\omega_{2}\rho^{2},
\end{equation}
where $p_{0}$, $\omega_{1}$ and $\omega_{2}$ are constants. Easily we can set $p_{0}=\omega_{2}=0$  to recover linear barotorpic equation of state.\\
Modified Chaplygin gas include only barotropic fluid with linear equation of state, while it is possible to extend them to including quadratic barotropic equation of state given by (\ref{I1}). It yields to introducing extended Chaplygin gas with the following equation of state \cite{P75 1402.2592,P Astro,P76 1405.0667,RP2014},
\begin{equation}\label{I7}
p_{ECG}=\sum{A_{n}\rho_{ECG}^{n}}-\frac{B}{\rho_{ECG}^{\alpha}}.
\end{equation}
It is obvious that the $n=1$ reduced to the modified Chaplygin gas. In this paper, we would focus on the second order term which recovers quadratic barotropic equation of state,
\begin{equation}\label{I8}
p_{ECG}=A_{1}\rho_{ECG}+A_{2}\rho_{ECG}^{2}-\frac{B}{\rho_{ECG}^{\alpha}},
\end{equation}
where $\alpha$, $A_{1}$, $A_{2}$ and $B$ are free parameters of the model.\\
On the other hand the temperature behavior and the thermodynamic stability of the generalized Chaplygin gas has been studied by the Ref. \cite{PLB636(2006)86}, and it is found that the generalized Chaplygin gas cools down through the expansion
without facing any critical point or phase transition. Also, thermodynamics of the generalized Chaplygin gas was investigated by introducing the integrability condition, and thermodynamicalquantities have been derived as functions of either the
volume or the temperature \cite{0812.0618}. Validity of the generalized second law of gravitational thermodynamics in a non-flat Friedmann-Robertson-Walker (FRW) universe and an expanding G\"odel-type universe containing the generalized Chaplygin gas was confirmed by the Refs. \cite{P26 1103.4842} and  \cite{IJTP52(2013)4583} respectively.\\
In extension of the Ref. \cite{PLB636(2006)86}, the similar work was performed for the case of the modified Chaplygin gas \cite{PLB646(2007)215} and the same result was obtained. More discussion on thermodynamical behavior of the modified Chaplygin gas can be found in the Ref. \cite{P58 1006.1461}. Also, Ref. \cite{P59 1012.5532} extended the work of \cite{0812.0618} to the case of the modified Chaplygin gas. Validity of the generalized second law of thermodynamics in the presence of the modified Chaplygin gas was investigated by the Ref. \cite{P60 1102.1632} and it was observed that the generalized second law of thermodynamics always was satisfied for the modified Chaplygin gas model. The generalized second law of thermodynamics in the brane-world scenario including the modified Chaplygin gas was verified for the late time behavior of the apparent horizon \cite{ASS341(2012)689}.\\
In this paper, we study a cosmological model where the universe is dominated by the extended Chaplygin gas. Indeed, we would like to investigate the evolution of density perturbations, trying to verify if the dominance of the extended Chaplygin gas is compatible with the formation of the universe. In that case we consider both Newtonian and relativistic descriptions of the extended Chaplygin gas. We find that the universe dominated by the extended Chaplygin gas is stable without any phase transition and critical point. We also investigate the thermodynamical aspects of the model and confirm the stability of the model using laws of thermodynamics.
\section{Extended Chaplygin gas cosmology}
\begin{figure}[h!]
 \begin{center}$
 \begin{array}{cccc}
\includegraphics[width=60 mm]{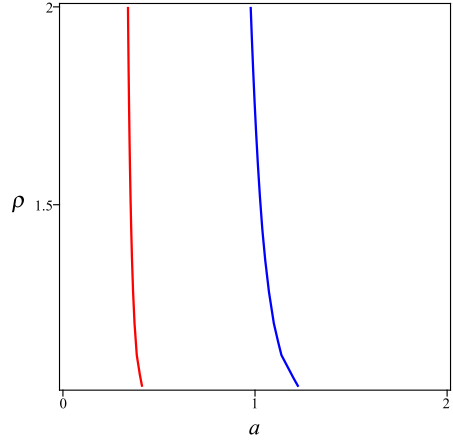}
 \end{array}$
 \end{center}
\caption{Typical behavior of $\rho$ in terms of $a$. blue: exact solution, red: approximate solution.}
 \label{fig0}
\end{figure}

Einstein's equations are given by,
\begin{equation}\label{s-1}
R_{\mu\nu}-\frac{1}{2}g_{\mu\nu}R=8\pi G T_{\mu\nu},
\end{equation}
where the energy momentum tensor is
\begin{equation}\label{s0}
T_{\mu\nu}=(p+\rho)u_{\mu}u_{\nu}-pg_{\mu\nu}.
\end{equation}
Assuming flat FRW universe,
\begin{equation}\label{s1}
ds^2=dt^2-a^{2}(dr^2+r^{2}(d\theta^{2}+\sin^{2}\theta d\phi^{2})),
\end{equation}
yield to the following Friedmann equations \cite{P93 1205.3768},
\begin{eqnarray}\label{s2}
H^{2}=\left(\frac{\dot{a}}{a}\right)^{2}=\frac{\rho_{ECG}}{3},\nonumber\\
\dot{H}=-\frac{1}{2}(p_{ECG}+\rho_{ECG}),
\end{eqnarray}
where $H$ is the Hubble expansion parameter and $a$ is the scale factor. Then conservation equation read as,
\begin{equation}\label{s3}
\dot{\rho}_{ECG}+3\frac{\dot{a}}{a}(p_{ECG}+\rho_{ECG})=0,
\end{equation}
where $8\pi G=1$ is used.
In order to simplify the calculations and to reduce the number of free parameters of the model and also to obtain an analytical expression of the energy density in terms of the scale factor, we assume the following conditions,
\begin{eqnarray}\label{s4}
\alpha&=&1,\nonumber\\
A_{1}&=&A_{2}-1,\nonumber\\
B&=&2A_{2}.
\end{eqnarray}
So, the equation of state (\ref{I8}) yields to the pure Chaplygin gas plus quadratic barotropic fluid equation of state. While it is very special case of the extended Chaplygin gas, we will show that the model may be suitable to describe universe. Having an explicit expression of the energy density in terms of the cosmological parameter is our motivation to choose the above condition.
Therefore, only free parameter of the model is $A_{2}$, and we can solve the conservation equation (\ref{s3}) using (\ref{I8}) and (\ref{s4}) to obtain the following relation,
\begin{equation}\label{s05}
\ln{a}=\frac{\ln{(\rho_{ECG}^{2}+2\rho_{ECG}+2)}}{30A_{2}}-\frac{\ln{(\rho_{ECG}-1)}}{15A_{2}}-\frac{\arctan{(\rho_{ECG}+1)}}{5A_{2}}+\mathcal{C},
\end{equation}
where $\mathcal{C}$ is an integration constant. In order to obtain an analytical solution we use $\tan^{-1}{(\rho_{ECG}+1)}\approx \pi/2$ approximation which is exact for for $\rho_{ECG}\ll1$. In the Fig. \ref{fig0} we show that exact and approximate solutions have similar manner and we can use approximate solution, so we can rewrite the equation (\ref{s05}) as follows,

\begin{equation}\label{s5}
\rho_{ECG}=1+\frac{2+\sqrt{-1+5a^{30A_{2}}e^{3\pi}}}{-1+a^{30A_{2}}e^{3\pi}e^{3\pi}},
\end{equation}
where $\tan^{-1}{(\rho_{ECG}+1)}\approx \pi/2$ approximation is used. This approximation is valid when $\rho_{ECG}\gg1$ corresponding to the early universe. However our solution will be valid at all times and our approximate solution is very close to the late time behavior with $\rho_{ECG}\ll1$. This is due to the fact that $\tan^{-1}{(\rho_{ECG}+1)}\approx \pi/4$, for $\rho_{ECG}\ll1$ therefore small compared to the logarithm term.\\
Hence, the Hubble parameter can written as follow,
\begin{equation}\label{s7}
H=\sqrt{\frac{1}{3}+\frac{2+\sqrt{-1+5a^{30A_{2}}e^{3\pi}}}{-3+3a^{30A_{2}}e^{3\pi}}}.
\end{equation}
Solving the above equation for $a$ gives the scale factor in terms of time. We can obtain the following relation,
\begin{equation}\label{s8}
t-f(a)+g(a)+C=0,
\end{equation}
where $C$ is an arbitrary integration constant, and we defined,
\begin{eqnarray}\label{s9}
f(a)&\equiv&\frac{\sqrt {3}\ln
 \left( 2\,\sqrt {-1+5\,{{\rm e}^{3\,\pi }}{a}^{30 A_{2}}}+1 \right)
}{15A_{2}},\nonumber\\
g(a)&\equiv&\frac{2\sqrt {3}\arctan \left( \sqrt {-1+5\,{{\rm e}^{3\,\pi
}}{a}^{30  A_{2}}} \right) }{15A_{2}}.
\end{eqnarray}
The contribution of the second function $g(a)$ is small compared to the logarithm term, so we can remove it and fix the constant $C$ to capture the effect of $g(a)$. Therefore, we can obtain the following time-dependence for the scale factor,
\begin{equation}\label{s10}
a=\frac{e^{-\frac{\pi}{10A_{2}}}}{20}[X(t)^{2}-2X(t)+5]^{\frac{1}{30A_{2}}},
\end{equation}
where we defined,
\begin{equation}\label{s11}
X(t)\equiv e^{5\sqrt{3}A_{2}(t+t_{0})},
\end{equation}
and $g(a)+C=t_{0}$ is used. It is clear that the scale factor is an increasing function of time.
\section{Relativistic analysis}
\begin{figure}[h!]
 \begin{center}$
 \begin{array}{cccc}
\includegraphics[width=60 mm]{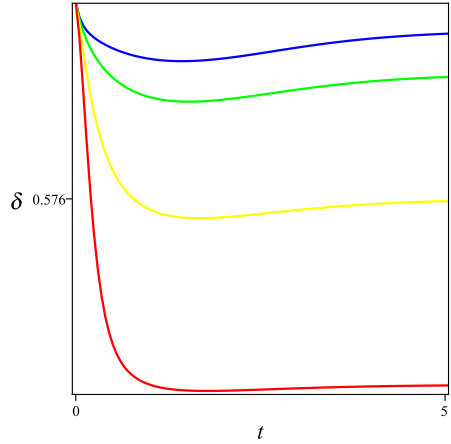}
 \end{array}$
 \end{center}
\caption{Typical behavior of $\delta$ in terms of $t$ for finite $A_{2}$ (extended Chaplygin gas). $n=0$ (blue), $n=1$ (green), $n=2$ (yellow), $n=3$ (red).}
 \label{fig1}
\end{figure}

\begin{figure}[h!]
 \begin{center}$
 \begin{array}{cccc}
\includegraphics[width=60 mm]{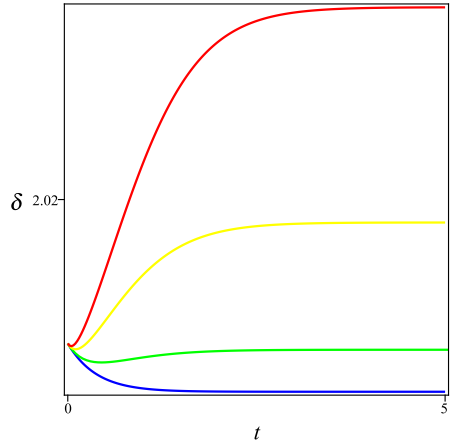}
 \end{array}$
 \end{center}
\caption{Typical behavior of $\delta$ in terms of $t$ for infinitesimal $A_{2}$ (modified Chaplygin gas). $n=0$ (blue), $n=1$ (green), $n=2$ (yellow), $n=3$ (red).}
 \label{fig2}
\end{figure}

We can study density perturbations of the Einstein's equations using the following relations,
\begin{eqnarray}\label{R1}
\tilde{g}_{\mu\nu}&=&g_{\mu\nu}+h_{\mu\nu},\nonumber\\
\tilde{\rho}&=&\rho+\delta\rho,\nonumber\\
\tilde{p}&=&p+\delta p,\nonumber\\
\tilde{u}_{\mu}&=&u_{\mu}+\delta u_{\mu},
\end{eqnarray}
where $h_{\mu\nu}, \delta\rho, \delta p$, and $\delta u_{\mu}$ are small perturbations around $g_{\mu\nu}, \rho, p$, and $u_{\mu}$ respectively. Assuming plane wave expansion to describe the spatial behavior of the perturbations and also the synchronous gauge condition $h_{\mu0}=0$ give us the following general equations \cite{GRG34(2002)53},
\begin{equation}\label{R2}
\ddot{h}+2H\dot{h}=\delta\rho+3\delta p,
\end{equation}
\begin{equation}\label{R3}
\dot{\delta}\rho+(p+\rho)\theta+3(p+\rho)\delta\rho-\frac{1}{2}(p+\rho)\dot{h}=0,
\end{equation}
\begin{equation}\label{R4}
(p+\rho)\dot{\theta}+(\dot{p}+\dot{\rho})\theta+5H(p+\rho)\theta=\frac{n^{2}}{a^{2}}\delta p,
\end{equation}
where $n$ is the wavelength of the perturbations, $\theta$ is velocity potential with definition $\nabla_i \theta = u_i$ \cite{P1 0003288} and $\delta=\delta\rho/\rho$. Also,
\begin{equation}\label{R5}
\delta p=\left(1-\frac{1}{A_{2}}+2\rho+\frac{2}{\rho^{2}}\right)A_{2}\delta\rho.
\end{equation}
We can obtain behavior of $\delta$ using simple assumption that $\dot{\theta}$ be a constant. In that case from the equations (\ref{R3}) and (\ref{R4}) we can obtain,
\begin{equation}\label{R6}
\theta=\frac{\frac{n^{2}}{a^{2}}(p\dot{\delta}+\delta \dot{p}-2H\delta p)-2c(\dot{p}+\dot{\rho})-5cH(p+\rho)}{\ddot{p}+\ddot{\rho}+5H(\dot{p}+\dot{\rho})+5\dot{H}(p+\rho)},
\end{equation}
\begin{equation}\label{R7}
\dot{h}=\frac{2}{p+\rho}[\rho\dot{\theta}+3\rho(p+\rho)\delta+(p+\rho)\theta],
\end{equation}
\begin{eqnarray}\label{R8}
\frac{(p+\rho)}{2}\ddot{h}&=&\rho\ddot{\delta}+\dot{\delta}\dot{\rho}+3\dot{\rho}(p+\rho)\delta+3\rho(\dot{p}+\dot{\rho})\delta+3\rho(p+\rho)\dot{\delta}\nonumber\\
&+&(p+\rho)c+(\dot{p}+\dot{\rho})\theta-\frac{\dot{p}+\dot{\rho}}{p+\rho}[\rho\dot{\delta}+3\rho(p+\rho)\delta+(p+\rho)\theta],
\end{eqnarray}
where $\dot{\theta}=c$ ($c$ is a constant) is used. Now, using the equations (\ref{R6}), (\ref{R7}), (\ref{R8}) in the equation (\ref{R2}) give us a second order differential equation for $\delta$. Note that using the equation (\ref{s10}) we can obtain $p$, $\rho$ and $H$ in terms of time. Resulting equation can be solved numerically to give us the behavior of $\delta$ as illustrated in the Fig. \ref{fig1}. We can see that perturbation may vanish at the late time for $n\approx3$. In order to compare the result with the previous versions such as modified Chaplygin gas we assume an infinitesimal value for $A_{2}$ and obtain results which are presented in the Fig. \ref{fig2}. We can see opposite behavior, so perturbations may vanish for $n=0$ and initially it behaves as $t^{m}$ ($m=2/3$ for dust), which agrees with the result of the Ref. \cite{GRG34(2002)53}.  As we know, in the limit of
small wavelength of the perturbations, the relativistic problem reduces to the
Newtonian problem. So, it is interesting to study the perturbations in Newtonian description which we will perform in next section.

\section{Newtonian description}
\begin{figure}[h!]
 \begin{center}$
 \begin{array}{cccc}
\includegraphics[width=60 mm]{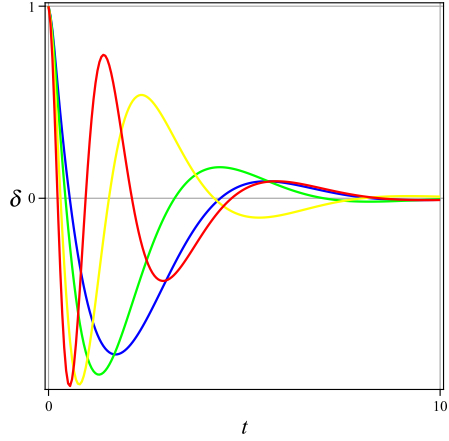}
 \end{array}$
 \end{center}
\caption{Typical behavior of $\delta$ in terms of $t$ for finite $A_{2}$ (extended Chaplygin gas). $n=0$ (blue), $n=1$ (green), $n=2$ (yellow), $n=3$ (red).}
 \label{fig3}
\end{figure}

The evolution equations for perturbation in Newtonian description are given by \cite{GRG34(2002)53},
\begin{equation}\label{N1}
\ddot{\delta}+2H\dot{\delta}+(\frac{v_{s}^{2}n^{2}}{a^{2}}+3\frac{\ddot{a}}{a})\delta=0,
\end{equation}
where,
\begin{equation}\label{N2}
v_{s}^{2}=\frac{\partial p}{\partial\rho}=1-\frac{1}{A_{2}}+2\rho+\frac{2}{\rho^{2}},
\end{equation}
is sound speed which is always positive for $A_{2}\geq1$ or $A_{2}<0$. Later we will give more details about the squared sound speed. In Fig. \ref{fig3} we can see that perturbations vanish at the late time and the model is completely stable without any phase transition and critical point.

\section{Thermodynamics}
\begin{figure}[h!]
 \begin{center}$
 \begin{array}{cccc}
\includegraphics[width=60 mm]{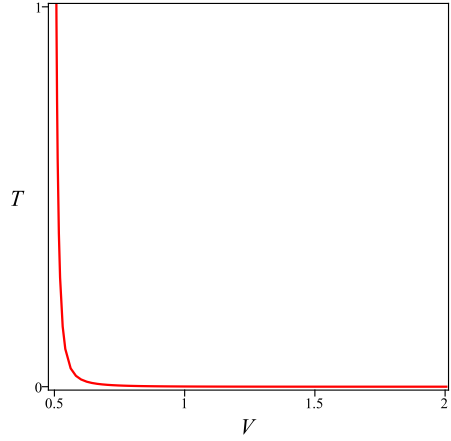}
 \end{array}$
 \end{center}
\caption{The temperature in terms of $V$ for $A_{2}=1.3$.}
 \label{fig4}
\end{figure}

\begin{figure}[h!]
 \begin{center}$
 \begin{array}{cccc}
\includegraphics[width=60 mm]{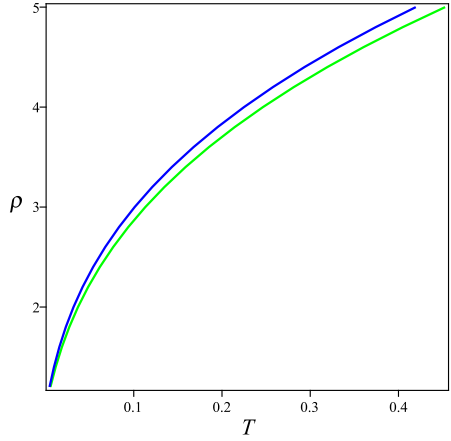}
 \end{array}$
 \end{center}
\caption{The energy density in terms of the temperature for $A_{2}=1.3$. Green line corresponds to the exact solution and blue line corresponds to the approximation solution.}
 \label{fig5}
\end{figure}

The beginning point of thermodynamical behavior of the model is the relation of some thermodynamical quantities with the energy density \cite{P59 1012.5532},
\begin{equation}\label{T1}
\rho=\frac{U}{V},
\end{equation}
where $U$ is the internal energy and $V=a^{3}$ is the volume of the universe. Another important quantity is the entropy which is given by,
\begin{equation}\label{T2}
S=\frac{A}{4G}.
\end{equation}
Using $8\pi G=1$ and $A=4\pi/H^{2}$ \cite{718583} gives us the following expression for entropy in terms of the Hubble expansion parameter,
\begin{equation}\label{T3}
S=8\pi^{2}H^{-2}.
\end{equation}
Using the relations (\ref{s2}) and  (\ref{s5}) we have Hubble expansion parameter in terms of $V$, which is decreasing function and yields to a constant as expected, therefore we can extract the entropy in terms of $V$ as follows,
\begin{equation}\label{T4}
S={\frac {24{\pi }^{2} \left( {V}^{10\,A_{{2}}}{{\rm e}^{3\,\pi }}-1
 \right) }{1+\sqrt {-1+5\,{V}^{10\,A_{{2}}}{{\rm e}^{3\,\pi }}}+{V}^{
10\,A_{{2}}}{{\rm e}^{3\,\pi }}}}.
\end{equation}
It is clear that the entropy is an increasing function of $V$ (and therefore is an increasing function of time) which tells that the generalized second law of thermodynamics is valid as expected. These are obtained by choosing $A_{2}>0$ and we usually fix $A_{2}=4/3$ to have $A_{1}=1/3$ and $B\approx3$ in agreement with some previous works such as \cite{P92 1106.4620}.\\
Then, using the first law and the second law of thermodynamics, the temperature can be written as \cite{P59 1012.5532},
\begin{equation}\label{T5}
T=\frac{(p+\rho)V}{S}.
\end{equation}
Assuming $V_{0}\approx0.5$ as initial volume of the universe gives the behavior of the temperature as illustrated in the Fig. \ref{fig4}. We can see high temperature at the early universe which decreases at the late time. It means that the temperature $T \gg0$ as $V \rightarrow V_{0}$ and $T \rightarrow 0$ as
$V \gg V_{0}$, therefore the third law of thermodynamics is satisfied for the extended Chaplygin gas model at the second order.\\
It is interesting to obtain a relationship between the temperature and the energy density. Using the $\tan^{-1}{(\rho_{ECG}+1)}\approx \pi/2$ approximation and above relations we can obtain,
\begin{equation}\label{T6}
T=\frac{A_{2}e^{-\frac{3\pi}{10A_{2}}}}{24\pi^{2}}(\rho^{3}+\rho^{2}-2)\left[\frac{\rho^{2}+2\rho+2}{\rho^{2}-2\rho+2}\right]^{\frac{1}{10A_{2}}},
\end{equation}
while the exact relation yields to the behavior represented in the Fig. \ref{fig5}. Again we can see a good agreement between approximate solution and exact solution. As we expected the energy density increases by increasing temperature, and we find $\rho\propto T^{1/3}$ at the early universe.\\

\begin{figure}[h!]
 \begin{center}$
 \begin{array}{cccc}
\includegraphics[width=60 mm]{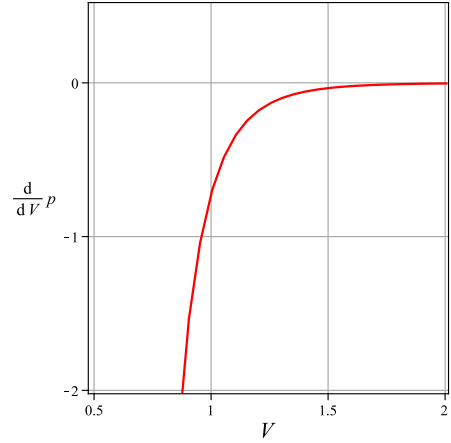}
 \end{array}$
 \end{center}
\caption{$\frac{\partial p}{\partial V}$ in terms of $V$ for $A_{2}=1.3$.}
 \label{fig6}
\end{figure}

\begin{figure}[h!]
 \begin{center}$
 \begin{array}{cccc}
\includegraphics[width=60 mm]{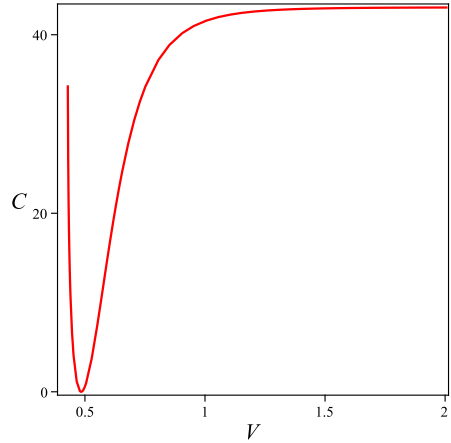}
 \end{array}$
 \end{center}
\caption{Heat capacity in terms of $V$ for $A_{2}=1.3$.}
 \label{fig7}
\end{figure}

In order to verify the thermodynamic stability conditions of the extended Chaplygin gas along its evolution, it is necessary
to have the following relations,
\begin{eqnarray}\label{T7}
\frac{\partial p}{\partial V}\leq0,\nonumber\\
T\frac{\partial S}{\partial T}>0.
\end{eqnarray}
The last equation is indeed the heat capacity which can be rewritten in terms of $V$ \cite{PLB646(2007)215},
\begin{equation}\label{T8}
C=V\frac{\partial\rho}{\partial V}\left(\frac{\partial T}{\partial V}\right)^{-1}.
\end{equation}
Figs. \ref{fig6} and \ref{fig7} show that both conditions in (\ref{T7}) are satisfied without any restriction. It means that the extended Chaplygin gas behaves as
a thermodynamically stable fluid (at least to the second order).

\begin{figure}[h!]
 \begin{center}$
 \begin{array}{cccc}
\includegraphics[width=50 mm]{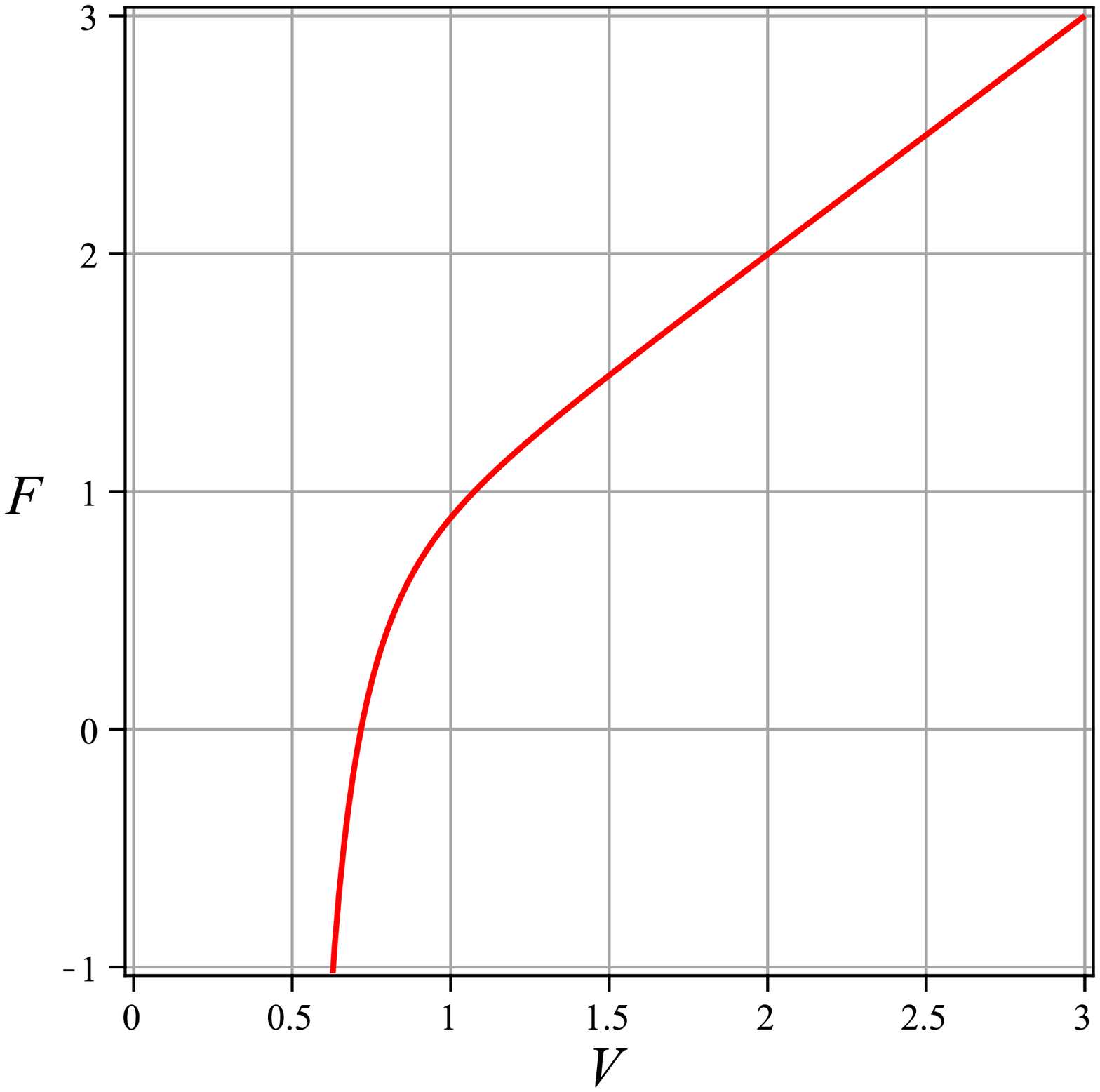}\includegraphics[width=50 mm]{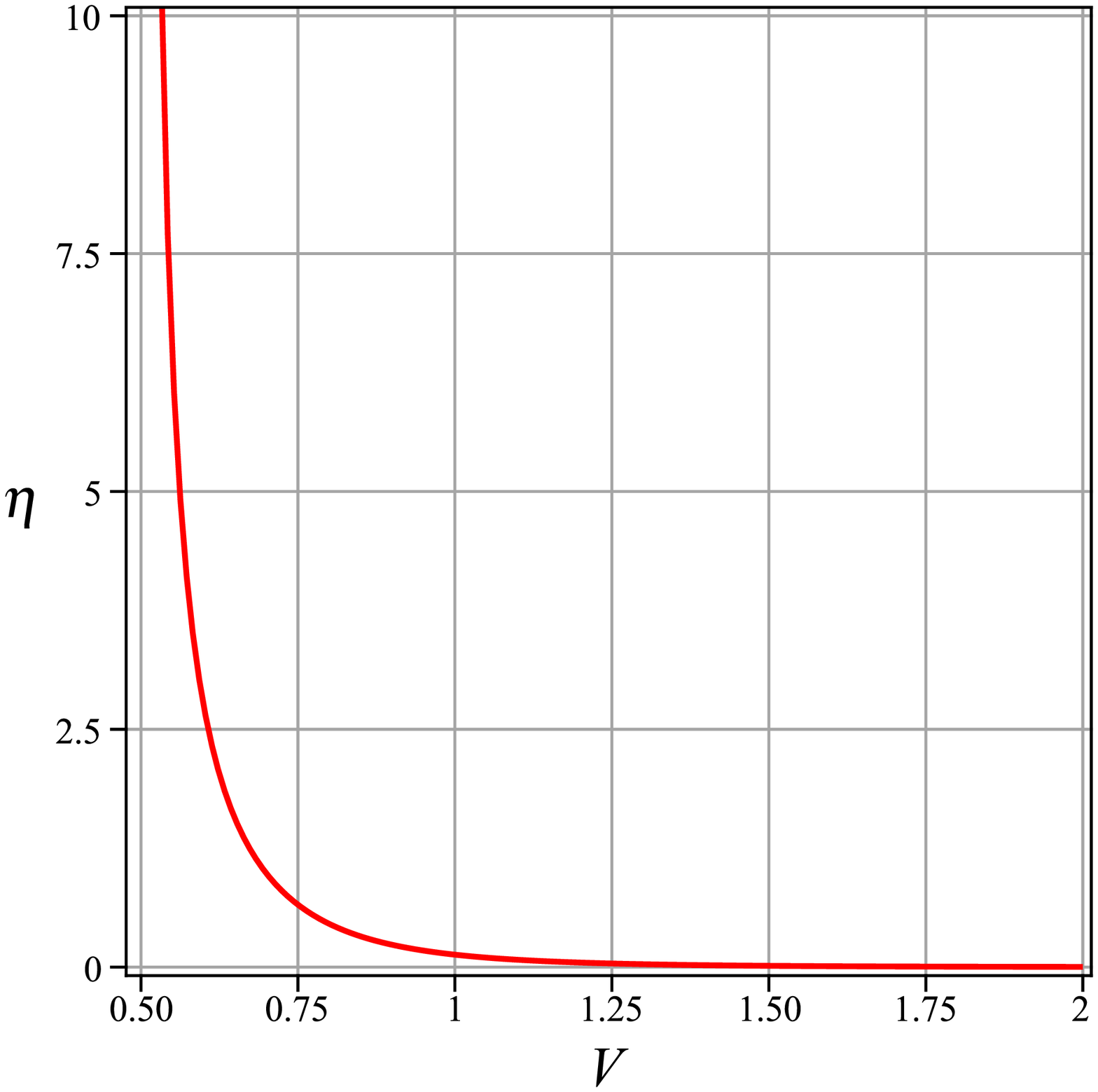}
 \end{array}$
 \end{center}
\caption{Free energy (left) and enthalpy (right) in terms of $V$ for $A_{2}=1.3$.}
 \label{fig8}
\end{figure}

In order to get the partition function, we need to calculate free energy which is given by,
\begin{equation}\label{T9}
F=U-TS,
\end{equation}
where internal energy $U$ is obtained using the equations (\ref{s5}) and (\ref{T1}). There is also another interesting quantity so called enthalpy (usually denoted by $H$ but we denote with $\eta$ to avoid conflict with Hubble parameter) which is given by,
\begin{equation}\label{T10}
\eta=U+pV.
\end{equation}
In the plots of the Fig. \ref{fig8} we represented the behavior of free energy and enthalpy in terms of $V$. We can see that the free energy is an increasing function of $V$, while enthalpy decreased to reach a constant at the late times. The enthalpy behaves as Hubble parameter with larger value at the initial time and smaller value at the late times. There is a critical $V$ (close to the present epoch) where $\eta=H$.

\begin{figure}[h!]
 \begin{center}$
 \begin{array}{cccc}
\includegraphics[width=60 mm]{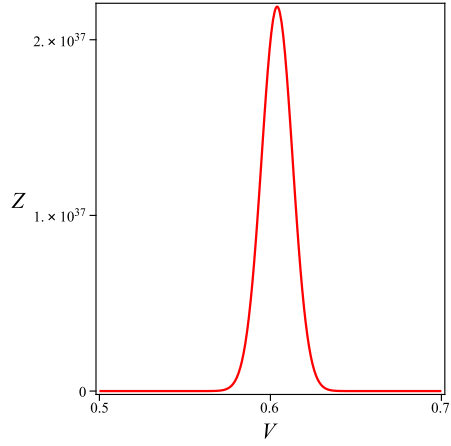}
 \end{array}$
 \end{center}
\caption{Partition function in terms of $V$ for $A_{2}=1.3$.}
 \label{fig9}
\end{figure}

Now, partition function can obtained using the following relation,
\begin{equation}\label{T11}
Z=e^{-\frac{F}{T}}.
\end{equation}
In the Fig. \ref{fig9}, we can see behavior of the partition function with $V$. The maximum value of $Z$ is near the early universe and it has small value at the present epoch. Before the extremum point the value of the free energy is negative, and the extremum point corresponds to zero free energy.

\section{Cosmological parameters}
\begin{figure}[h!]
 \begin{center}$
 \begin{array}{cccc}
\includegraphics[width=50 mm]{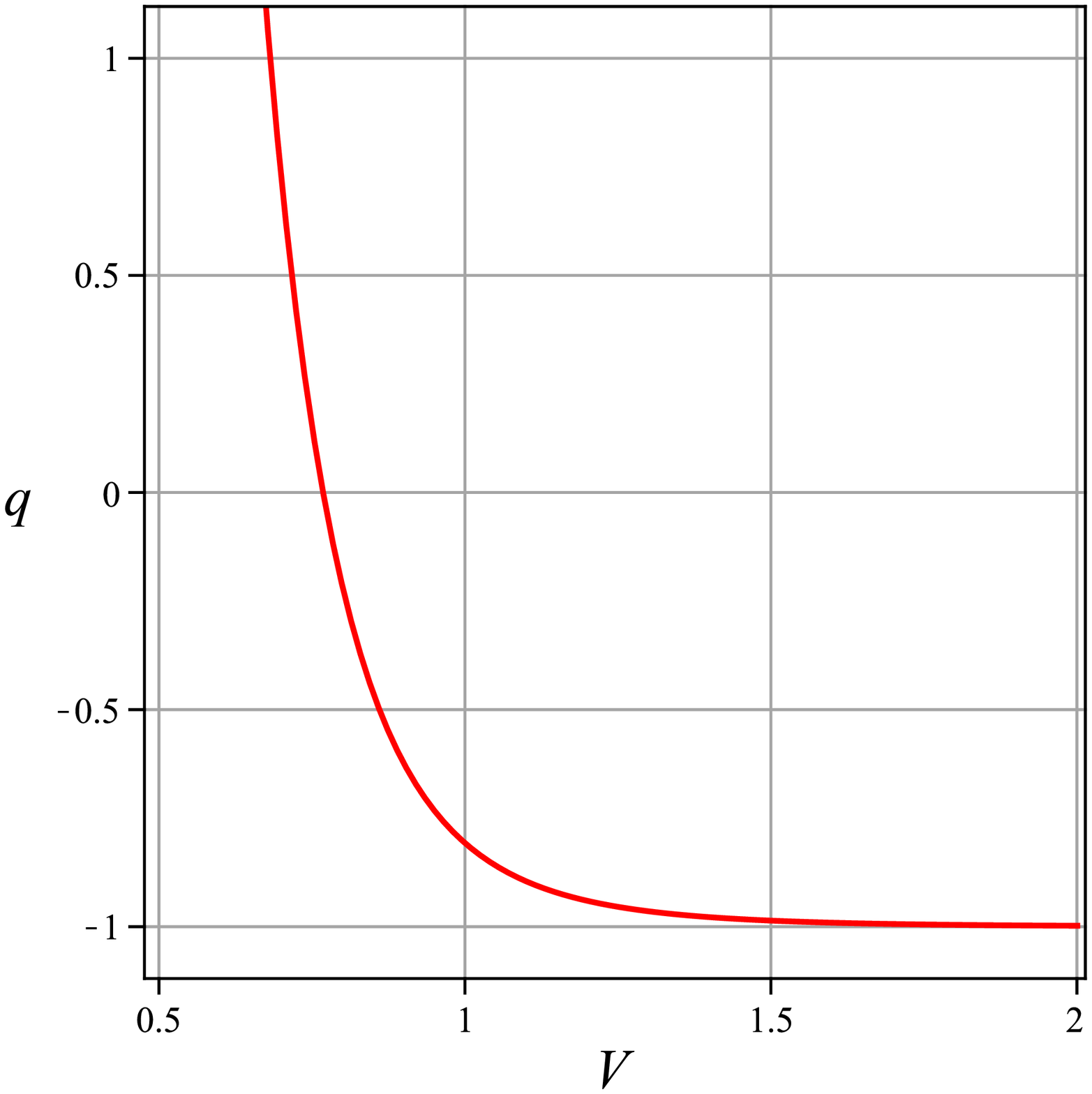}\includegraphics[width=50 mm]{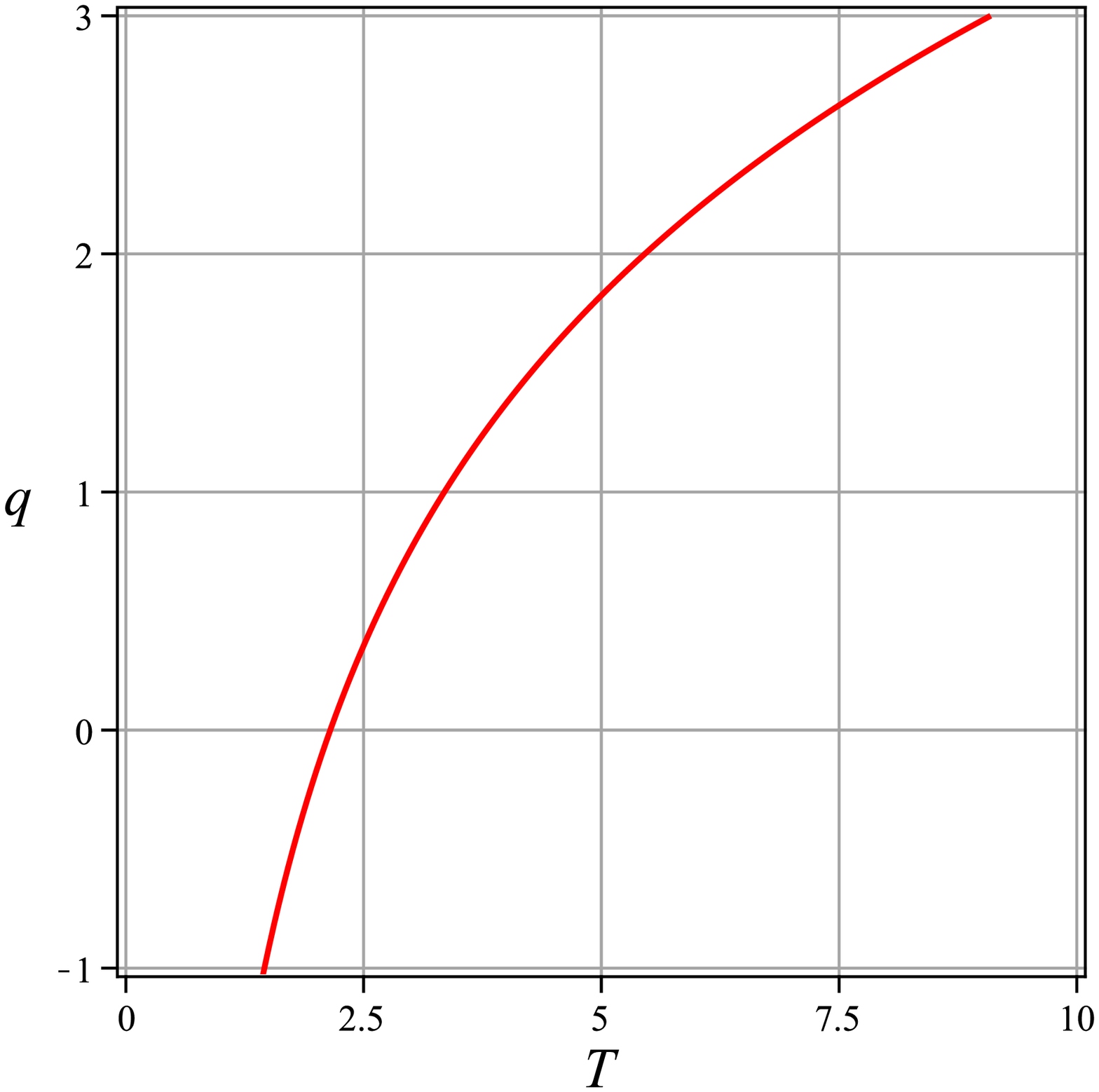}
 \end{array}$
 \end{center}
\caption{Deceleration parameter in terms of $V$ (left) and temperature (right) for $A_{2}=1.3$.}
 \label{fig10}
\end{figure}

\begin{figure}[h!]
 \begin{center}$
 \begin{array}{cccc}
\includegraphics[width=50 mm]{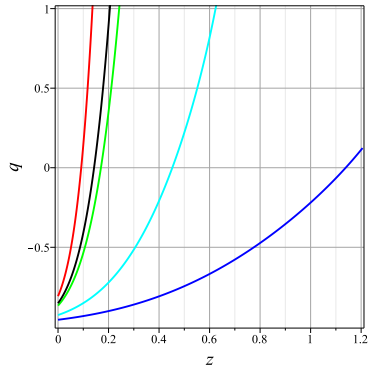}
 \end{array}$
 \end{center}
\caption{Deceleration parameter in terms of redshift for $A_{2}=0.3$ (blue), $A_{2}=0.5$ (cyan), $A_{2}=0.9$ (green), $A_{2}=1$ (black) and $A_{2}=1.3$ (red).}
 \label{fig101}
\end{figure}

We can investigate some cosmological parameters by varying thermodynamical quantities. An important parameter is the deceleration parameter given by,
\begin{equation}\label{C1}
q=-1+\frac{p+\rho}{2H^{2}}.
\end{equation}
Therefore we can study the behavior of the deceleration parameter in terms of $V$ and $T$ (see Fig. \ref{fig10}). As expected $q\rightarrow-1$ is obtained at the late times. Also, deceleration to acceleration phase transition happens in this model. We give a plot of the deceleration parameter in terms of redshift by Fig. \ref{fig101} to see at what $z$ the late acceleration begins for various possible values of $A_{2}$. For instance, in the case of $A_{2}=1.3$ we see that universe starts accelerating at $z\leq0.09$.

\begin{figure}[h!]
 \begin{center}$
 \begin{array}{cccc}
\includegraphics[width=60 mm]{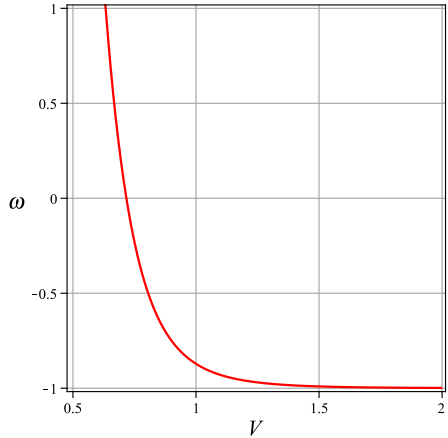}
 \end{array}$
 \end{center}
\caption{$\omega$ in terms of $V$ for $A_{2}=1.3$.}
 \label{fig11}
\end{figure}

The behaviour of the equation of state parameter $\omega=p/\rho$ is also illustrated in the Fig. \ref{fig11}. As expected it yields to -1 at the late times. Some time at the early universe we have $\omega=0$ so the extended Chaplygin gas behaves as dust ($p=0$) and exchanges to de sitter at the late times.

\begin{figure}[h!]
 \begin{center}$
 \begin{array}{cccc}
\includegraphics[width=60 mm]{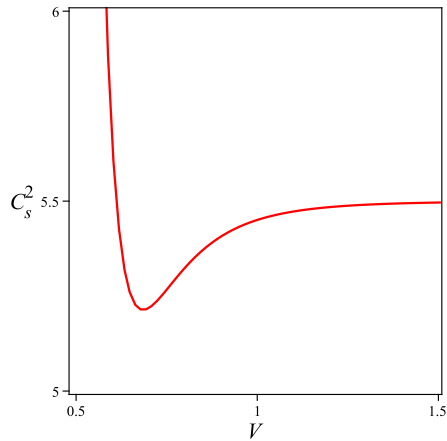}
 \end{array}$
 \end{center}
\caption{squared sound speed in terms of $V$ for $A_{2}=1.3$.}
 \label{fig12}
\end{figure}

Finally, we can investigate the squared sound speed (as promised before) which is an important parameter to study the stability of the model. As we know the fluid will be stable if,
\begin{equation}\label{C2}
v_{s}^{2}=\frac{\partial p}{\partial\rho}\geq0.
\end{equation}
In Fig. \ref{fig12} we choose $A_{2}=1.3$ and see that $C_{s}^{2}\geq0$ without any restriction, therefore there is no phase transition and the model will be stable at all times.\\
So we demonstrated the stability of the ECG model using two separate ways, analyzing our model cosmologically and thermodynamically. Agreement of cosmological parameters with observational data supports our model.

\section{Conclusion}
In this paper, we considered extended Chaplygin gas at the second order which recover quadratic barotropic equation of state. First of all we obtained energy density in terms of scale factor and found an analytical expression for the scale factor. Then, we studied the evolution of density perturbations in both relativistic and Newtonian regime. In the relativistic regime we have shown that the perturbations may vanish for higher wavelength of perturbations. This is completely opposite with the modified Chaplygin gas where perturbations may vanish at small wavelength of perturbations. At the Newtonian limit we have shown that the perturbations vanish without any limitation. We also studied thermodynamical quantities of the model such as temperature and entropy. We confirmed that the thermodynamics laws satisfied and fluid is thermodynamically stable all times. Stability of the fluid also is investigated by using the sound speed. We found that there is no phase transition and critical point in this model.  Partition function is also calculated in the model using the free energy. Using the numerical analysis we found that the free energy is an increasing function of time (also $V$). Analyzing of enthalpy yields to an interesting result that is similar to the Hubble parameter at about present epoch. We investigated the behavior of some important cosmological parameters in terms of thermodynamical quantities. For example $q\rightarrow-1$ and $\omega\rightarrow-1$ was shown to have desired behaviours at late times. We can also consider higher order terms and expect to obtain similar results.

\end{document}